\documentclass[10pt,twocolumn]{IEEEtran}
\usepackage{graphicx}
\usepackage{verbatim}

 \def\theenumi{\hspace{5mm}\roman{enumi}}

\begin{document}

\title{\Large\bf Active Internet Traffic Filtering: 
Real-time Response to Denial-of-Service Attacks}

\author{Katerina Argyraki \hspace{1cm} David R. Cheriton\\
{\it Computer Systems Lab\\
Stanford University\\
\{argyraki, cheriton\}@dsg.stanford.edu}\\
\hspace{1cm} \\
Subj-class: Networking and Internet Architecture}

\date{}
\maketitle

\thispagestyle{empty}

\renewcommand{\labelitemi}{$-$}

\begin{abstract}
Denial of Service (DoS) attacks are one of the most challenging threats to 
Internet security.
An attacker typically compromises a large number of vulnerable hosts and uses them
to flood the victim's site with malicious traffic,
clogging its tail circuit and interfering with normal traffic.
At present, the network operator of a site under attack has no other resolution
but to respond manually
by inserting filters in the appropriate edge routers to drop attack traffic.
However, as DoS attacks become increasingly sophisticated,
manual filter propagation becomes unacceptably slow or even infeasible.

In this paper, we present \emph{Active Internet Traffic Filtering},
a new automatic filter propagation protocol.
We argue that this system provides a guaranteed, significant level of
protection against DoS attacks
in exchange for a reasonable, bounded amount of router resources.
We also argue that the proposed system cannot be abused by a malicious
node to interfere with normal Internet operation.
Finally, we argue that it retains its efficiency in the face of continued
Internet growth.
\end{abstract}

\section{Introduction}
\label{intro}

Denial of Service (DoS) attacks are recognized as one of the most 
challenging threats to Internet security.
Any organization or enterprise that is dependent on the Internet
can be subject to a DoS attack,
causing its service to be severely disrupted, if not fail completely.
The attacker typically uses a worm to create an ``army'' of zombies,
which she orchestrates to flood the victim's site with malicious traffic.
This malicious traffic exhausts the victim's resources,
thereby seriously affecting the victim's ability to respond to normal traffic.

A network layer solution is required because the end-user or end-organization
has no way to protect its tail circuit from being congested by an attack,
causing the disruption sought by the attacker.
For example, if an enterprise has a $10$ Mbps connection to the Internet,
an attacker can command its zombies to send traffic far exceeding this $10$ Mbps 
rate to this enterprise,
completely congesting the downstream link to the enterprise
and causing normal traffic to be dropped.

Network operators use conventional router filtering capabilities to respond to DoS attacks.
Typically, an operator of a site under attack identifies the nature of the
packets being used in the attack by some packet collection facility,
installs a filter in its firewall/edge router to block these packets
and then requests its ISP to install comparable filters in its routers
to remove this traffic from the tail circuit to the site.
Each ISP can further communicate with its peering ISPs to block this unwanted
traffic as well, if it so desires.

Currently, this propagation of filters is manual:
the operator on each site determines the necessary filters
and adds them to each router configuration.
In several attacks,
the operators of different networks have been forced to communicate by
telephone given that the network connection, and thus email, was inoperable
because of the attack.

As DoS attacks become increasingly sophisticated,
manual filter propagation becomes unacceptably slow or even infeasible.
For example, an attack can switch from one protocol to another,
move between source networks as well as oscillate between on and off
far faster than any human can respond.
In general, network operators are confronting an ``arms race'' in which
any defense, such as manually installed filters,
is viewed as a challenge by the community of attacker-types to defeat.
Exploiting a weakness such as human speeds of filter configuration
is an obvious direction for an attacker to pursue.

The concept of automatic filter propagation has already been introduced in 
\cite{Mahajan01}: a router is configured with a filter to drop (or rate-limit) 
certain traffic; if it continues to drop a significant amount of this traffic,
it requests that the upstream router take over and block the traffic.
However, the crucial issues associated with automatic filter propagation
are still unaddressed.

The real problem is how to efficiently manage the bounded number of
filters available to a network operator to provide this filtering support.
An attacker can change protocols, source addresses, port numbers, etc.
requiring a very large number of filters.
However, a sophisticated hardware router has a fixed maximum number of
wire-speed filters that can block traffic with no degradation
in router performance.
The maximum is determined by hardware table sizes
and is typically limited to several thousand.
A software router is typically less constrained by table space,
but incurs a processing overhead for each additional filter.
This usually limits the practical number of filters to even less than
a hardware router.
Moreover, there is a processing cost at each router for installing each new
filter, removing the old filters and sending and receiving filter
propagation protocol messages.

Given the restricted amount of filtering resources available to each router,
hop-by-hop filter propagation towards the attacker's site clearly does not scale:
Internet backbone routers would quickly become the ``filtering bottleneck''
having to satisfy filtering requests coming from all the corners of the Internet.
Fortunately, traceback \cite{Savage00} \cite{Snoeren01} 
makes it possible to identify a router close to the attacker
and send it a filtering request directly.
However, any filter propagation mechanism other than hop-by-hop raises a serious
security issue:
Once a router starts accepting filtering requests from unknown sources,
how can it trust that these requests are not forged by malicious nodes
seeking to disrupt normal communication between other nodes?

In this paper we propose a new filter propagation protocol called AITF
(Active Internet Traffic Filtering): 
The victim sends a filtering request to its network gateway.
The victim's gateway temporarily blocks the undesired traffic, 
while it propagates the request to the attacker's gateway.
As we will see,
the protocol both motivates and assists the attacker's gateway to block
the attack.
Moreover, a router receiving a filtering request satisfies it
only if it determines that the requestor is on the same path with
the specified undesired traffic.
Thus, the filter cannot affect any nodes in the Internet
other than those already operating at the mercy of the requestor.

The novel aspect of AITF is that it enables each participating
service provider to guarantee to its clients a \emph{specific, significant amount
of protection against DoS attacks},
while it requires only a \emph{bounded credible amount of resources}.
At the same time it is \emph{secure} i.e., it cannot be abused by a malicious 
node to harm (e.g. block legitimate traffic to) other nodes.
Finally, it \emph{scales with Internet size} i.e., it keeps its efficiency
in the face of continued Internet growth.

\section{Active Internet Traffic Filtering (AITF)}
\label{aitf}

\subsection{Terminology}

A \emph{flow label} is a set of values that captures the common characteristics
of a traffic flow --
e.g., ``all packets with IP source address $S$ and IP destination address $D$''.

A \emph{filtering request} is a request to block a flow of packets --
all packets matching a specific wildcarded flow label -- for the next
$T$ time units.

A \emph{filtering contract} between networks $A$ and $B$ specifies:
\begin{enumerate}
\item{The filtering request rate $R_{1}$ at which $A$ accepts filtering
requests to block certain traffic to $B$.}
\item{The filtering request rate $R_{2}$ at which $A$ can
send filtering requests to get $B$ to block certain traffic from coming into $A$.}
\end{enumerate}

An \emph{AITF network} is an Autonomous Domain which has a filtering contract
with each of its end-hosts and each neighbor Autonomous Domain directly connected to it.
An \emph{AITF node} is either an end-host or a border router 
\footnote{A \emph{border router} is a router that has interfaces in more
than one AITF networks.}
in an AITF network.

Finally, we define the following terms with respect to an undesired flow:
The \emph{attack path} is the set of AITF nodes the undesired flow goes through.
The \emph{attacker} is the origin of the undesired flow.
The \emph{victim} is the target of the undesired flow.
The \emph{attacker's gateway} is the AITF node closest to the attacker
along the attack path.
Similarly, the \emph{victim's gateway} is the AITF node closest to the victim
along the attack path.

\subsection{Overview}
\label{overview}

The AITF protocol enables a service provider to protect a client
against $N$ undesired flows, 
by using only $n\ll N$ filters and a DRAM cache of size $O(N)$.
The motivation is that each router can afford gigabytes of DRAM
but only a limited number of filters. 

In an AITF world, 
each Autonomous Domain (AD) is an AITF network
i.e., it has filtering contracts with all its end-hosts and peering ADs.
These contracts limit the rates by which the AD can send/receive
filtering requests to/from its end-hosts and peering ADs.
The limited rates allow the receiving router to police the requests to the specified rates
and indiscriminately drop requests when the rate is in excess of the agreed rate.
Thus, the router can limit the CPU cycles used to process filtering requests
as well as the number of filters it requires.

An AITF filtering request is initially sent from the victim to 
the victim's gateway;
the victim's gateway propagates it to the attacker's gateway;
finally, the attacker's gateway propagates it to the attacker.
Both the victim's gateway and the attacker's gateway install filters to block 
the undesired flow.
The victim's gateway installs a filter only temporarily, to immediately
protect the victim, while it waits for the attacker's gateway to take responsibility.
The attacker's gateway is expected to install a filter and block the undesired flow for
$T$ time units.

If the undesired flow stops within some grace period,
the victim's gateway interprets this as a hint that the attacker's gateway 
has taken over and removes its temporary filter.
This leaves the door open to ``on-off'' undesired flows
\footnote{When the attacker's gateway does not cooperate,
the attacker can start an undesired flow, stop long enough to trick the victim's gateway into
removing its temporary filter, then start again and so on.}.
In order to detect and block such ``on-off'' flows, 
the victim's gateway needs to remember each
filtering request for at least $T$ time units.
Thus, the victim's gateway, installs a filter for $T_{tmp} \ll T$ time units,
but keeps a ``shadow'' of the filter in DRAM for $T$ time units
\footnote{
Keeping each filter for $T$ time units is very expensive --
doing so would defeat the whole purpose of pushing filtering to the attacker's gateway.}.

The attacker's gateway expects the attacker to stop the undesired flow within a grace period.
Otherwise, it holds the right to disconnect from her.
This fact encourages the attacker to stop the undesired flow.
Similarly, the victim's gateway expects the attacker's gateway to block the 
undesired flow within a grace period.
Otherwise, the mechanism \emph{escalates}: 
The victim's gateway now plays the role of the victim 
(i.e., it sends a filtering request to its own gateway)
and the attacker's gateway plays the role of the attacker
(i.e., it is asked to stop the undesired flow or risk disconnection).
The escalation process should become clear with the example in
\ref{example}.

Thus, the mechanism proceeds in \emph{rounds}.
At each round, only four nodes are involved.
In the first round, 
the mechanism tries to push filtering of undesired traffic back to 
the AITF node closest to the attacker.
If that fails, it tries the second closest AITF node to the attacker and so on.

\subsection{Basic protocol}
\label{protocol}

The AITF protocol involves only one type of message: a \emph{filtering request}. 
A filtering request contains a \emph{flow label} and a \emph{type} field. 
The latter specifies whether this request is addressed to the victim's gateway,
the attacker's gateway or the attacker.

The only nodes in an AITF network that speak the AITF protocol are end-hosts
and border routers. Internal routers do not participate.

AITF node $X$ sends a filtering request to AITF node $Y$,
when $X$ wants a certain traffic flow coming through $Y$ to be blocked
for $T$ time units.

When AITF node $Y$ receives a filtering request, 
it checks which end-host or peering network the request is received
from/through.
If that end-host or peering network has exceeded its allowed rate,
the request is dropped.
If not, $Y$ looks at the specified undesired flow label and takes certain actions:
\begin{itemize}
\item{If $Y$ is the victim's gateway: 
\begin{enumerate}
\item{It installs a \emph{temporary} filter to block the undesired flow for 
$T_{tmp} \ll T$ time units.}
\item{It logs the filtering request in DRAM for $T$ time units.}
\item{It propagates the filtering request to the attacker's gateway.
If the attacker's gateway does not block the flow within $T_{tmp}$ time units,
$Y$ propagates the filtering request to its own gateway.}
\end{enumerate} 
}
\item{If $Y$ is the attacker's gateway: 
\begin{enumerate}
\item{It installs a filter to block the undesired flow for $T$ time units.}
\item{It propagates the filtering request to the attacker.
If the attacker does not stop the flow within a grace period,
$Y$ disconnects from her.}
\end{enumerate}
}
\item{If $Y$ itself is the attacker, it stops the flow (to avoid disconnection).}
\end{itemize}

We should note that the behavior described above is that of a non-compromised,
non-malicious node.
Neither the attacker not even the attacker's gateway
are expected to always conform to this behavior.
AITF operation does not rely on their cooperation.

\subsection{Example}
\label{example}

\begin{figure}[htbp]
\begin{center}
\resizebox{7cm}{!}{\includegraphics{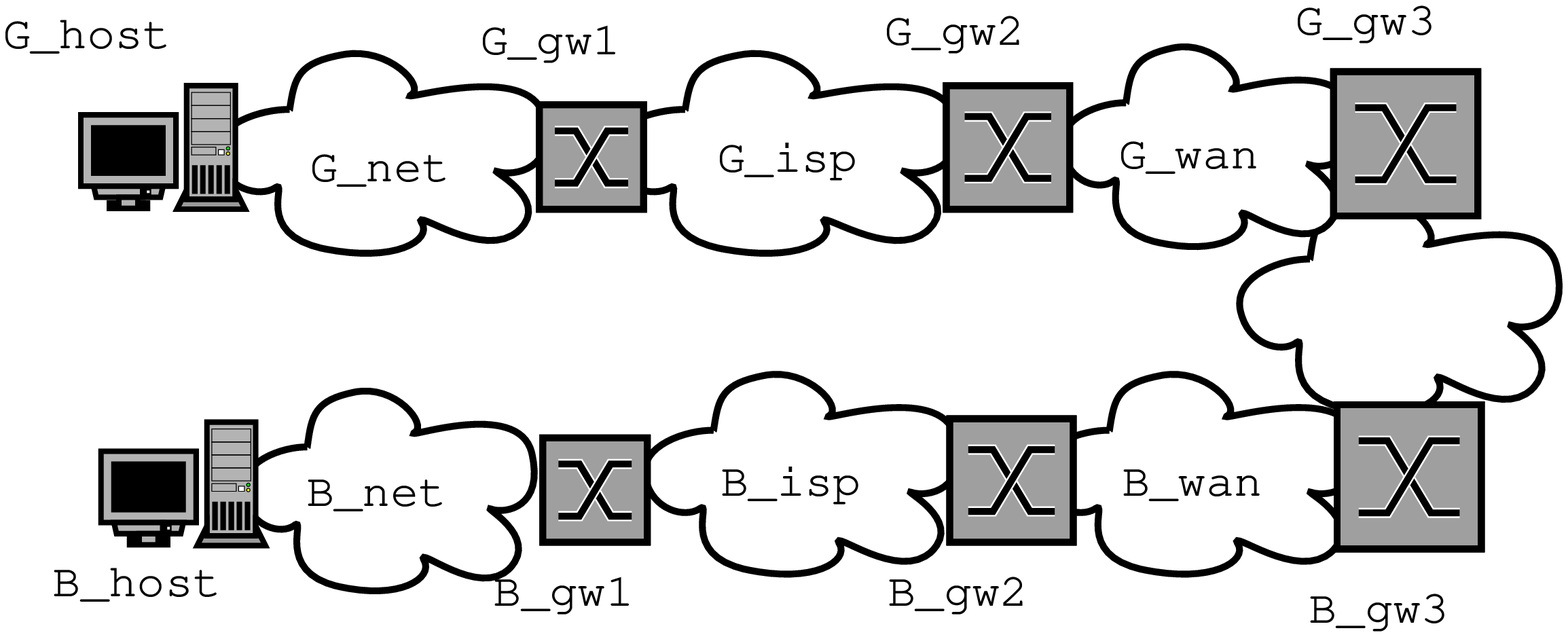}}
\caption{Example attack path from attacker $B\_host$ to victim $G\_host$}
\label{filterfig}
\end{center}
\end{figure}

In Figure \ref{filterfig} $G\_host$ -- which stands for ``good host'' -- is an end-host 
residing in enterprise network $G\_net$, 
which is connected to local ISP $G\_isp$ through router $G\_gw1$.
$G\_isp$ runs a regional network that connects through its backbone router 
$G\_gw2$ to a wide-area ISP $G\_wan$.
Similarly, $B\_host$ -- which stands for ``bad host'' -- is an end-host residing in
enterprise network $B\_net$ etc.

$B\_host$ starts sending an undesired flow to $G\_host$.
$G\_host$ sends a filtering request to $G\_gw1$ against $B\_host$. 
Upon reception of $G\_host$'s request, $G\_gw1$ temporarily blocks
the undesired flow but also propagates the request to $B\_gw1$.

On the other side, upon reception of $G\_gw1$'s request,
$B\_gw1$ immediately blocks the undesired flow,
but also propagates the filtering request to $B\_host$. 
$B\_host$ either stops the undesired flow or risks being disconnected.
Thus, if $B\_gw1$ cooperates, 
by the end of the first round,
filtering of the undesired flow has been successfully pushed to the AITF
node closest to the attacker ($B\_gw1$).

Of course, $B\_gw1$ may decide not to cooperate and ignore the filtering request.
Then, the mechanism escalates:
$G\_gw1$ propagates the filtering request to $G\_gw2$.
$G\_gw2$ temporarily blocks all undesired traffic,
but also propagates the filtering request to $B\_gw2$ and so on.
Thus, if $B\_gw2$ cooperates, 
by the end of the second round,
filtering of the undesired flow has been successfully pushed to the 
second closest to the attacker AITF node ($B\_gw2$).

In the worst-case scenario, even $B\_gw3$ refuses to cooperate.
As a result, $G\_gw3$ disconnects from $B\_gw3$.

\subsection{Verifying a filtering request}
\label{handshake}

In a network architecture where source address spoofing is allowed,
compromised node $M$ can maliciously request the blocking of traffic
from $A$ to $V$ thereby disrupting their communication.
To avoid this, we add a simple extension to the basic protocol.

The extension introduces two more messages:
A \emph{verification query} and a \emph{verification reply}.
Both types include a \emph{flow label} and a \emph{nonce} (i.e., a random number).

When router $Y$ receives a filtering request,
which asks for the blocking of a traffic flow from attacker $A$ to victim $V$,
$Y$ verifies that the request is real before taking any action to satisfy it.
If $Y$ is the victim's gateway,
this verification is trivial with appropriate ingress filtering.
If $Y$ is the attacker's gateway,
verification is accomplished through the following ``3-way handshake'':

\begin{enumerate}
\item{Router $Y$ receives a filtering request,
asking for the blocking of a traffic flow from attacker $A$ to victim $V$.}
\item{$Y$ sends a verification query to $V$,
asking ``Do you really not want this traffic flow?''}
\item{$V$ responds to $Y$ with a verification reply.
The reply must include the same flow label and nonce included in the query.
}
\end{enumerate}
If the nonce on $V$'s reply is the same with the nonce on $Y$'s query,
$Y$ accepts the request as real and proceeds to satisfy it.
The ``3-way handshake'' is further discussed in \ref{secure}.

\subsection{Assumptions}
\label{assumptions}

AITF operation assumes that the victim's gateway can determine
\begin{itemize}
\item{Who is the attacker's gateway (in order to propagate the request).}
\item{Who is the next AITF node on the attack path (in order to escalate, if necessary).} 
\end{itemize}
These assumptions are met, if an efficient traceback technique as those described in
\cite{Savage00} \cite{Snoeren01} is available.

Also AITF assumes that off-path traffic monitoring is not possible
i.e., if node $M$ is not located on the path from node $A$ to node $V$,
then $M$ cannot monitor traffic sent on that path
(this assumption is necessary for the ``3-way handshake'').

\section{Discussion}
\label{discussion}

\subsection{Why it works}
\label{work}

The basic idea of AITF is to push filtering of undesired traffic to
the network closest to the attacker.
That is, hold a service provider responsible for providing connectivity to a 
misbehaving client and have it do the dirty job.
The question is, why would the attacker's service provider accept 
(or at least be encouraged) to do that?

If the attacker's service provider does not cooperate, 
it risks being disconnected by its own service provider.
This makes sense for both of them:
If $B\_net$ in Figure \ref{filterfig} refuses to block its misbehaving client,
the filtering burden falls on $B\_isp$. 
Thus, it makes sense for $B\_isp$ to consider $B\_net$ a bad client and disconnect from it. 
On the other hand, this offers an incentive to $B\_net$ to cooperate 
and block the undesired flow.
Otherwise, it will be disconnected by $B\_isp$,
which will result in all of its clients being dissatisfied.

Moreover, AITF offers an economic incentive to providers to protect their network 
from the inside by employing appropriate ingress filtering.
If a provider pro-actively prevents spoofed flows from exiting its network,
it lowers the probability of an attack being launched from its own network,
thus reducing the number of expected filtering requests it will later have
to satisfy to avoid disconnection.

In short, AITF creates a cost vs quality trade-off for service providers:
Either they pay the cost to block the undesired flows generated by their few bad clients,
or they run the risk of dissatisfying their legitimate clients, 
which are the vast majority.
Thus, the quality of a provider's service is now related to its capability to filter its
own misbehaving clients.

\subsection{Why it is secure}
\label{secure}

The greatest challenge with automatic filtering mechanisms is that 
compromised node $M$ may maliciously request the blocking of traffic from $A$ to $V$,
thereby disrupting their communication.
AITF prevents this through the ``3-way handshake'' described in \ref{handshake}.

The ``3-way handshake'' does not exactly verify the authenticity of a filtering request.
It only enables $A$'s gateway to verify that a request to block traffic 
from $A$ to $V$ has been sent by a node located on the path from $A$ to $V$.
A compromised router located on this path can naturally forge and snoop 
handshake messages to disrupt $A$-$V$ communication.
However, such a compromised router can disrupt $A$-$V$ communication anyway,
by simply dropping the corresponding packets.
\footnote{We assumed that off-path traffic monitoring is impossible.
Thus, an off-path malicious node cannot snoop handshake messages.} 

In short, AITF cannot be abused by a compromised node to cause interruption
of a legitimate traffic flow,
unless that compromised node is responsible for routing the flow,
in which case it can interrupt the flow anyway.

\subsection{Why it scales}
\label{scale}

AITF scales with Internet size, 
because it pushes filtering of undesired traffic to the leaves of the Internet, 
where filtering capacity follows Internet growth.

In most cases,
AITF pushes filtering of undesired traffic to the provider(s) of the attacker(s).
Thus, the amount of filtering requests a provider is asked to satisfy
grows proportionally to the number of the provider's (misbehaving) clients.
However, intuitively, a provider's filtering capacity also grows proportionally
to the number of its clients.
\footnote{In the AITF model, the filtering contract becomes part of a provider's services.
Thus, a part of each client's service fee is invested in provisioning the provider's routers
with extra filters.}
In short, a provider's filtering capacity follows the provider's filtering workload.

If the attacker's provider is itself compromised,
AITF naturally fails to push filtering to it.
Instead, filtering is performed by another network, closer to the Internet core.
If this situation occurred often, then the scalability argument stated above would be false.
Fortunately,
compromised routers are a very small percentage of the Internet infrastructure. 
\footnote{It is difficult to compromise a router, because of the limited number of 
potentially exploitable services it offers.
Moreover, service providers have an immense economic incentive to keep their routers
uncompromised.} 
Thus, AITF fails to push filtering to the attacker's provider with a very small
probability.

\section{Performance Analysis}
\label{analysis}

In this section we provide simple formulas that describe AITF performance.
For lack of space and given that our formulas are very simple and intuitive, 
we defer any details to \cite{aitf}.

\subsection{The victim's perspective}
\label{victim}

\subsubsection{Effective bandwidth of an undesired flow}

AITF significantly reduces the \emph{effective bandwidth} of an undesired flow
-- i.e., the bandwidth of the undesired flow actually experienced by the victim.
Specifically, it can be shown that AITF reduces the effective bandwidth of an 
undesired flow by a factor of
$$r \approx \frac{n(T_d + T_r)}{T}$$
where $n$ is the number of non-cooperating 
\footnote{
By \emph{non-cooperating} node we mean an AITF node, which does not take
its responsibility to filter an undesired flow for $T$ time units.
}
AITF nodes on the attack path,
$T_d$ is attack detection time and $T_r$ is the one-way delay from the
victim to its gateway. 
$T$ is the timeout associated with all filtering requests
i.e. each filtering request asks for the blocking of a flow for $T$ time units.
For example, if the only non-cooperating node on the attack path is the attacker,
and if the one-way delay from the victim to its gateway is $T_r = 50$ msec,
for $T = 1$ min,
an AITF node can reduce the effective bandwidth of an undesired flow by
a factor $r \approx 0.00083$.
\footnote{
$T_d$ may be significant the first time the undesired flow is detected.
Here, we ignore that initial overhead.
Detecting a reappearing undesired flow could be as fast as matching a received
packet header to a logged undesired flow label i.e., insignificant compared
to the one-way delay to the victim's gateway.
}

Here we only demonstrate this result for $n=1$ i.e.,
when the only non-cooperating node on the attack path is the attacker:
At time $0$ the attacker starts the undesired flow;
at time $T_d$ the victim detects it and sends a filtering request;
at time $T_d + T_r$ the victim's gateway temporarily blocks the flow
and the victim stops receiving it;
the flow is eventually blocked by the attacker's gateway and released after
time $T$.
Thus, if the original bandwidth of the undesired flow is $B$,
its effective bandwidth is $B_e \approx B\cdot\frac{T_d + T_r}{T}$.

When $n \ge 1$ i.e., when $1$ or more AITF routers close to the attacker
are non-cooperating, the attacker can play ``on-off'' games: 
Pretend to stop the undesired flow to trick the victim's gateway into removing
its filter, then resume the flow etc.
The victim's gateway detects and blocks such attackers by using its DRAM cache.

\subsubsection{Number of undesired flows}

An AITF node is guaranteed protection against a specific number of undesired flows,
which depends on its contract with its service provider.
Specifically, it can be shown that 
if a client is allowed to send $R_{1}$ filtering requests per time unit 
to the provider, then the client is protected 
\footnote{
When we say that the client is ``protected'' against an undesired flow,
we mean that the client can significantly reduce the effective bandwidth of the flow,
as described in the previous subsection.}
against 
$$N_v = R_{1}\cdot T$$ 
simultaneous undesired flows.
For example, for $R_1 = 100$ filtering requests per second and $T = 1$ min,
the client is protected against $N_v = 6,000$ simultaneous undesired flows.

\subsection{Filtering close to the victim}
\label{victims_gw}

AITF enables a service provider to protect a client against $N_v$ undesired
flows by using only $n_v \ll N_v$ filters.
Specifically, it can be shown that
if a client is allowed to send $R_{1}$ filtering requests per time unit to the provider,
the provider needs $n_v$ filters and a DRAM cache that can fit $m_v$ filtering requests
in order to satisfy all the requests,
where
$$n_v = R_{1}\cdot T_{tmp}, \ \ \ m_v = R_{1}\cdot T$$
$T_{tmp}$ is the amount of time that elapses from the moment the victim's gateway
installs a temporary filter until it removes it.
The purpose of the temporary filter is to block the undesired flow until
the attacker's gateway takes over.
Therefore, $T_{tmp}$ should be large enough to allow
the traceback from the victim's gateway to the attacker's gateway
plus the 3-way handshake. 
For example, suppose we use an architecture like \cite{CheritonGritter00},
where traceback is automatically provided inside each packet.
Then traceback time is $0$.
If the 3-way handshake between the two gateways takes $600$ msec,
for $R_1 = 100$ filtering requests per second and $T = 1$ min,
the service provider needs $n_v = 60$ filters to protect a client
against $N_v = 6,000$ undesired flows.

\subsection{Filtering close to the attacker}
\label{attackers_gw}

AITF requires a bounded amount of resources from the attacker's service provider.
Specifically,
if a service provider is allowed to send $R_{2}$ filtering requests 
per time unit to a client,
then the provider needs $$n_a = R_{2}\cdot T$$ filters in order to ensure that the client
satisfies all the requests.
Given these resources, the provider can filter $N_a = n_a = R_{2}\cdot T$ 
simultaneous undesired flows generated by the client.
For example, for $R_2 = 1$ filtering request per second and $T = 1$ min,
the provider needs $n_a = 60$ filters for the client.
This filtering request rate allows the provider to filter up to $N_a = 60$ 
simultaneous undesired flows generated by the client.

\subsection{The attacker's perspective}
\label{attacker}

We have defined an attacker as the source of an undesired flow.
By this definition, an attacker is not necessarily a malicious/compromised node;
it is simply a node being asked to stop sending a certain flow.
A legitimate AITF node must be provisioned to stop sending undesired flows 
when requested, in order to avoid disconnection.

AITF requires a bounded amount of resources from the attacker as well.
Specifically,
if a service provider is allowed to send $R_{2}$ filtering requests
per time unit to a client,
the client needs $$n_a = R_{2}\cdot T$$ filters 
(as many as the provider) in order to satisfy all the requests.
For example, for $R_2 = 1$ filtering request per second and $T = 1$ min,
the client needs $n_a = 60$ filters.

\section{Related Work}
\label{related}

In \cite{Mahajan01} Mahajan {\it et al.} propose mechanisms for detecting and
controlling high bandwidth traffic aggregates.
One part of their work discusses how a node determines whether it is congested
and how it identifies the aggregate(s) responsible for the congestion.
In contrast, we start from the point where the node has identified the undesired flow(s).
In that sense, their work and our work are complementary.
Another part of their work discusses how much to rate-limit an annoying aggregate
due to a DoS attack or a flash crowd.
In contrast,
our mechanism focuses on DoS attack traffic and attempts to limit it to rate $0$.
\footnote{
We believe that DoS attacks should be addressed separately from flash crowds:
Flash crowd aggregates are created by legitimate traffic.
Therefore, it makes sense to rate-limit them instead of completely blocking them.
On the contrary, DoS attack traffic aims at disrupting the victim's operation.
Therefore, it makes sense to block it.
Blocking a traffic flow is simpler and cheaper than rate-limiting it.
Moreover, DoS attack traffic is generated by malicious/compromised nodes.
Therefore, it demands a more intelligent defense mechanism.
}

The part of their work most related to ours proposes a cooperative {\it pushback} mechanism:
A congested node attempts to rate-limit an aggregate by dropping a portion of its packets.
If the drop rate remains high for several seconds,
the node considers that it has failed to rate-limit the aggregate and
asks its adjacent upstream routers to do it.
If the recipient routers also fail to rate-limit the aggregate,
they recursively propagate pushback further upstream.

A pushback request is propagated hop by hop by the victim towards the attacker.
In contrast, the propagation of an AITF filtering request involves only $4$ nodes:
the victim, the victim's gateway, the attacker's gateway and the attacker --
we claim that this allows AITF to scale with Internet size.
A pushback request does not force the recipient router to rate-limit the problematic
aggregate; it relies on its good will to cooperate.
In contrast,
AITF forces the attacker to discontinue the undesired flow
and the attacker's service provider to filter the attacker
or else risk disconnection --
we claim that this makes AITF deployable.

In \cite{ParkLee01} Park and Lee propose DPF (Distributed Packet Filtering),
a distributed ingress-filtering mechanism for pro-actively blocking spoofed flows.
In contrast, AITF aims at blocking \emph{all} undesired -- including spoofed -- flows
as close as possible to their sources. Thus, it cannot be replaced by DPF.
On the other hand, DPF blocks most spoofed flows \emph{before} they reach their destination
i.e., DPF is proactive, whereas AITF is reactive.
In that sense, DPF and AITF are complementary.

In \cite{Keromytis02} Keromytis {\it et al.} propose SOS (Secure Overlay Services),
an architecture for pro-actively protecting against DoS attacks 
the communication between a pre-determined location
and a specific set of users who have authorized
access to communicate with that location. 
In contrast, AITF addresses the more general problem of protecting 
against DoS attacks any location accessible to all Internet users.

Finally, \cite{Savage00} and \cite{Snoeren01}
propose traceback solutions i.e., mechanisms that enable a victim to 
reconstruct the path followed by attack packets in the face of source address
spoofing.
As already mentioned, an efficient traceback mechanism is necessary to AITF 
operation.

\section{Conclusions}
\label{conclusions}

We presented AITF, an automatic filter propagation mechanism,
according to which each Autonomous Domain (AD) has a filtering contract with each
of its end-hosts and neighbor ADs.
A filtering contract with a neighbor provides a guaranteed, significant level of 
protection against DoS attacks coming through that neighbor
in exchange for a reasonable, bounded amount of router resources.

Specifically:
\begin{itemize}
\item{Given a filtering contract between a client and a service provider,
which allows the client to send $R_1$ filtering requests per time unit to the provider,
the provider can protect the client against a large number of undesired flows 
$N_v = R_1 \cdot T$, 
by significantly limiting the effective bandwidth of each undesired flow. 
The provider achieves this by using only a modest number of filters 
$n_v = R_1 \cdot T_{tmp} \ll N_v$.}
\item{Given a filtering contract between a client and a service provider,
which allows the provider to send $R_2$ filtering requests per time unit to the client,
both the client and the provider need a bounded number of filters $n_a = R_2 \cdot T$
to honor their contract.}
\end{itemize}

We argued that AITF successfully deals with the biggest challenge 
to automatic filtering mechanisms: source address spoofing.
Namely, we argued that it is not possible for any malicious/compromised node 
to abuse AITF in order to interrupt a legitimate traffic flow,
unless the compromised node is responsible for routing that flow,
in which case it can interrupt the flow anyway.

Finally, we argued that AITF scales with Internet size, 
because it pushes filtering of undesired traffic to
the service providers of the attackers,
unless the service providers are themselves compromised.
Fortunately, compromised routers are a very small percentage of Internet 
infrastructure.
Thus, in the vast majority of cases, AITF pushes filtering of undesired traffic to the 
leaves of the Internet, where filtering capacity follows Internet growth.


\end{document}